# Artificial Intelligence Advances for De Novo Molecular Structure Modeling in Cryo-EM

Dong Si*[1], Andrew Nakamura[1], Runbang Tang[2], Haowen Guan[3], Jie Hou[4], Ammaar Firozi[4], Renzhi Cao[5], Kyle Hippe[5], Minglei Zhao[6]

[1]Division of Computing and Software Systems, University of Washington Bothell, Bothell, WA 98011

[2]Molecular Engineering and Sciences Institute, University of Washington Seattle, Seattle, WA 98105

[3]Applied and Computational Math Sciences, University of Washington Seattle, Seattle, WA 98195

[4]Department of Computer Science, Saint Louis University, Saint. Louis, MO 63103

[5]Department of Computer Science, Pacific Lutheran University, Tacoma, WA 98447

[6]Department of Biochemistry and Molecular Biology, University of Chicago, Chicago, IL 60637

*For correspondence: dongsi@uw.edu

## Abstract

Cryo-electron microscopy (cryo-EM) has become a major experimental technique to determine the structures of large protein complexes and molecular assemblies, as evidenced by the 2017 Nobel Prize. Although cryo-EM has been drastically improved to generate high-resolution three-dimensional (3D) maps that contain detailed structural information about macromolecules, the computational methods for using the data to automatically build structure models are lagging far behind. The traditional cryo-EM model building approach is template-based homology modeling. Manual de novo modeling is very time-consuming when no template model is found in the database. In recent years, de novo cryo-EM modeling using machine learning (ML) and deep learning (DL) has ranked among the top-performing methods in macromolecular structure modeling. Deep-learning-based de novo cryo-EM modeling is an important application of artificial intelligence, with impressive results and great potential for the next generation of molecular biomedicine. Accordingly, we systematically review the representative ML/DL-based de novo cryo-EM modeling methods. And their significances are discussed from both practical and methodological viewpoints. We also briefly describe the background of cryo-EM data processing workflow. Overall, this review provides an introductory guide to modern research on artificial intelligence (AI) for de novo molecular structure modeling and future directions in this emerging field.

## 1. Introduction

Cryo-electron microscopy (cryo-EM) began with the introduction of a specimen preparation method by Dubochet and coworkers in the 1980s[1]. One of the technologies, known as single-particle analysis, now allows researchers to resolve 3D maps of macromolecules at near-atomic resolution. The resolution of single-particle analysis depends on the signal-to-noise ratio of individual particle projections, the accuracy of 3D Euler angles determined through iterative reconstruction, the completeness of Fourier space sampling, and the conformational and compositional homogeneity of the sample. With the improved hardwares such as the direct electron detector (DED) and new image processing methods, cryo-EM has become a viable tool to analyze the structures of biomolecules, showing comparable performance to X-ray crystallography[2,3]. The process of single-particle analysis in cryo-EM can be described in three steps: sample preparation, data acquisition, and image processing. Particle picking is a crucial first step in the computational pipeline of image processing. It used to be a tedious and time-consuming step. However, this processing is extremely important as neglecting false-positive particles may lead to pitfalls known as "Einstein from noise"[4]. With the development of high-throughput data collection hardware and methods for single-particle analysis, a typical dataset contains millions of particles; thus, it is essential to use software tools to facilitate the process, and machine learning (ML) algorithms have a great application on this problem which have led to explosive growth of cryo-EM maps.

In 2003, the Worldwide Protein Data Bank (PDB) was founded to archive cryo-EM maps and structure models derived from cryo-EM maps[5]. Advances in cryo-EM technology and image analysis software started a 'resolution revolution' in 2013[6]. Subsequent hardware and software advancements further improved cryo-EM to achieve atomic resolutions of individual atoms in a protein. Direct electron detectors were used for 70% of structures in 2016, in contrast to none before 2011, indicating the widespread biomedical and pharmaceutical industry interest[7]. The widespread use of the cryo-EM equipment allowed research labs to transition from X-ray crystallography techniques to cryo-EM for modeling the 3D shapes of protein structures. In 2020, the data in EMDataResource (EMDR), previously known as EMDataBank (EMDB), has increased exponentially due to the growth in cryo-EM labs worldwide as well as due to the global crisis to derive structures that can fight the COVID-19 and future pandemics[8–18]. Researchers have also made use of methods that automatically and accurately determine the molecular structure from cryo-EM maps[19]. However, there are still manual modeling steps that prevent PDB EM models from reaching the rate at which new maps are deposited to EMDR/EMDB (**Figure 1**).

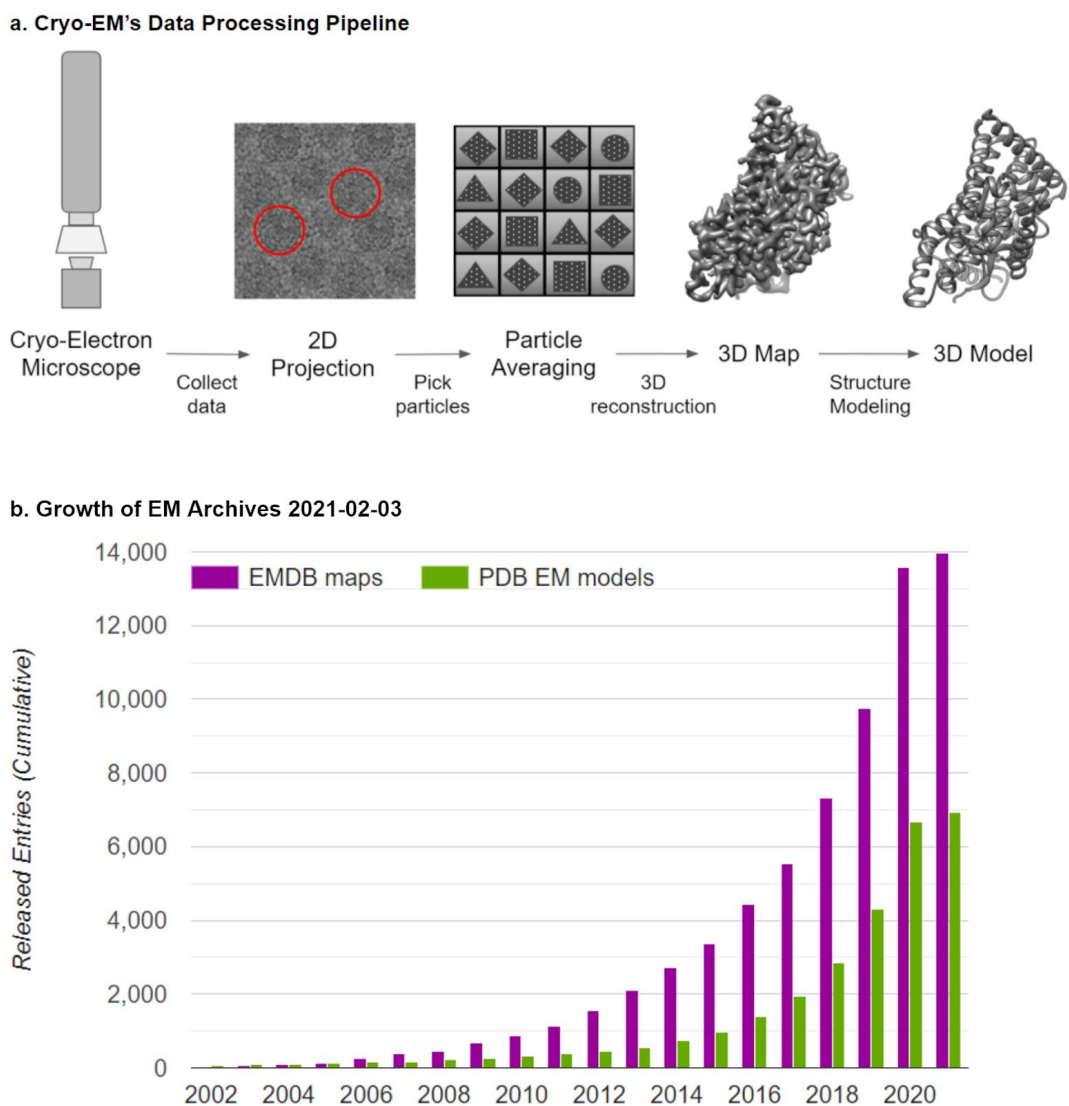

**Figure 1**. The process and growth of Cryo-EM. a. The cryo-EM data processing pipeline. 2D projections of sample molecules (particles) are visualized in electron micrographs. When cryo-EM captures the native state of the molecules, particles with different conformation can be found in the same data set. This then allows software to utilize the data for higher resolution 3D reconstruction and structure modeling. b. Growth of Electron Microscopy Archives[20]. In 2003, the Worldwide Protein Data Bank (PDB) was founded to archive PDB EM models and EMDB maps. In 2013, direct electron detectors improved and allowed for near-atomic resolution by 2016. In 2020, the EMDB has increased exponentially due

to the growth in cryo-EM microscopes in labs worldwide as well as due to the global crisis to derive structures that can fight the COVID-19 and future pandemics. Manual modeling steps prevent PDB EM models from reaching the rate at which new EMDB maps are submitted.

Over the past decades, the cryo-EM imaging technique has revolutionized the quality of 3D cryo-EM maps from the medium-resolution (5–10 Å) to near-atomic resolution (1.2–5 Å), even significantly higher in the foreseeable future. The cryo-EM maps reveal detailed structural information about specific macromolecules such as proteins. Several studies have shown that the secondary structure features (e.g., α-helix and β-sheets) can be captured from the medium-resolution maps[21–25]. Furthermore, backbone atoms and side-chain details are distinguishable at higher-resolution maps using advanced algorithms[19,26–28]. Therefore, the wealth of structural information in cryo-EM maps leads to many computational methods that have been implemented to derive the high-resolution protein structures from the cryo-EM maps. The most common approaches for building high-resolution structures from the intermediate-resolution map (~ 4-10 Å) include the rigid-fitting[24,29–55] and flexible-fitting[56–88] into cryo-EM maps[32,33,37,43,44,89–91], which consists of several major steps: predicting the atomic structures of the subunits, defining satisfactory goodness-of-fit between the atomic structure and the cryo-EM maps according to predefined scoring functions, searching the optimal location in the cryo-EM map for the atomic structures to fit, and performing flexible refinement to improve the structural fitting. It is worth noting that these steps have also been effectively integrated to assist the de novo protein structure prediction from the cryo-EM maps using optimization methods[92]. However, the computational approaches were limited to the low resolution of cryo-EM maps and the availability of homologous templates. If there is a nearly identical match to a target sequence, a previously determined 3D structure from the protein sharing a homologous sequence can be used to fit into the cryo-EM maps. If no homologs are found, de novo modeling can be used to generate an initial 3D structure from the monomer sequence alone. The recent application of deep learning (DL) techniques has greatly advanced the de novo protein structure prediction using the co-evolutionary sequence data[93–96]. For protein complex structure prediction, several homology modeling approaches have been developed to get high-resolution protein complex structures[97–100]. Despite the significant progress made in protein structure prediction from sequence, there are still challenges in automated de novo protein complex structure modeling that conforms to the 3D cryo-EM map.

Earlier cryo-EM structure modeling (model building) approaches often involve substantial human intervention, which can be labor-intensive and time-consuming. To address this challenge, advanced ML algorithms start being applied to each step of the process to achieve automation, thus relieving scientists from the laborious work and helping improve the efficiency and effectiveness of cryo-EM modeling. The pattern recognition driven by ML techniques has largely improved the direct identification of structural details from the 3D cryo-EM maps, ranging from the secondary structure elements (i.e., α-helix, β-sheets, coil)[21–25] to the backbone atoms or side-chain positions[19,26–28,101].

In this review, we first introduce several computational methods for single-particle picking in cryo-EM imaging and 3D reconstruction, in which several ML algorithms and DL techniques have been applied. The following sections discuss the development of de novo structure modeling methods from the cryo-EM maps using artificial intelligence, ranging from the identification of low-resolution structural elements using traditional ML methods to modeling high-resolution structure models using advanced DL techniques. We also review the recent advances for RNA/DNA structure modeling in cryo-EM maps that can promote the study of protein nucleic-acid interactions involved in biological processes. Finally, we discuss quality assessment, molecular dynamics and function prediction problems, and the potential trends in artificial intelligence based cryo-EM data mining and modeling.

# 2. AI in cryo-EM image data processing

## 2.1. Particle picking

The computational methods for automated particle picking from cryo-EM images aim to facilitate the 3D map reconstruction in single-particle data analysis. Several ML methods have been proposed. TextonSVM is a tool that uses a support vector machine algorithm (SVM) as the classification algorithm to recognize particles based on their texture[102]. Automatic Particle Picking with Low user Effort (APPLE) particle picker also uses SVM but takes a different approach by classifying particles and noise regions based on their different mean intensity and variance[103]. The recent advancement in DL has demonstrated better performance in large-scale data analysis than traditional ML approaches in particle picking. For instance, the convolutional neural network (CNN) is one of the DL algorithms that has shown outstanding performance in the field of computer vision and has also been applied in selecting particles from the cryo-EM images, including DeepPicker[104], FastParticlePicker[105], DeepEM[106], PIXER[107], Topaz[108], and crYOLO[109]. DeepPicker takes advantage of the image recognizing power of CNN to capture common features of particles from previously-analyzed micrographs and use the knowledge to locate the centers of particles. FastParticlePicker has a similar major component to DeepPicker. It applies the new "Fast Region-based CNN" framework that drastically reduces the runtime and achieves full automation with the help of cross-molecule training scheme. DeepEM can distinguish "good particles" from "bad" ones, including overlapped particles, local aggregates, background noise fluctuations, ice contamination, and carbon-rich areas, through the effort of learning the identity of particles with data augmentation (i.e., rotate the training data). PIXER is the first that applied the segmentation network concept in combination with CNN. It achieves full automation while also  acquires a result as good as DeepEM. Topaz frames particle picking as a positive-unlabeled (PU) learning problem, assisted with CNN combined with Autoencoder. The method eases researchers from the burden of labeling micrographs when it is hard to achieve. BoxNet[110] is a tool in the software package *Warp* for particle picking based on Residual Neural Network (ResNet) and it achieved a unique feature that can mask out high-contrast artifacts in micrographs when bundled with *Warp*. Besides the generally known ML and DL algorithms, AutoCryoPicker[111] proposed a new unsupervised clustering algorithm called intensity-based clustering (IBC). Using IBC, AutoCryoPicker achieved a fully automated single-particle picking that does not require any manual work even during the training process, such as labeling training data. In the test of comparing IBC with two traditional clustering algorithms K-means and Fuzzy C-means (FCM) and another particle picking software - EMAN2[112], IBC achieved the highest accuracy and also the fastest runtime. The ML-based particle picking methods are summarized in **Table 1**.

**Table 1**. Strength and limitation of particle picking methods.

| Method | ML/DL Algorithm | Strengths | Limitations | Tool |
|---|---|---|---|---|
| TextonSVM[102] (Aebeláez et al., 2011) | SVM | First particle picking algorithm that utilized ML; "Precision–recall" performance is significantly better than the previous correlation based method SIGNATURE[113] | Prone to human bias; Performance is not as good as humans | N/A |
| APPLE Particle picker[103] (Heimowitz et al., 2018) | SVM | Fully automated, uses no templates and requires no manual selection; Not vulnerable to manual or template bias | Imposes an assumption on the particle size, which may cause regions containing artifacts and can be | https://github.com /PrincetonUnivers ity/APPLEpicker |

| | | | mistaken for particle projections | |
|---|---|---|---|---|
| DeepPicker[104] (Wang et al., 2016) | CNN | Results are comparable to manual selection by human experts; Robust against a wide range of contrast/defocus levels of micrograph | Semi-automated, requires manual selection of (several hundreds) examples for training; Prone to manual bias; Uses a fixed threshold for particle selection; The precision-recall performance is ~ 60%-80% in the test | http://https//github.com/nejyeah/DeepPicker-python |
| FastParticlePicker[105] (Xiao and Yang, 2017) | Fast R-CNN | Fully automated; Drastically reduced the runtime without precision or recall loss | The precision-recall performance is similar to DeepPicker (~ 60%-80%); Vulnerable to low signal-to-noise (SNR) images | https://github.com/xiao1fan/FastParticlePicker |
| DeepEM[106] (Zhu et al., 2017) | CNN | Could distinguish unwanted particles; Uses a varied threshold; The precision-recall performance is ~90% on KLH dataset and ~80% on cryo-EM datasets | Semi-automated, requires manual selection of (several hundreds) examples for training | N/A |
| BoxNet[110] (Tegunov and Cramer, 2018) | Residual Neural Network (ResNet) | Took the advantage of ResNet and trained with a very deep model; Able to mask out high-contrast artifacts in cryo-EM images | Requires examples to do retraining for achieving the best performance; Particle picking performance was not reported in the paper | https://github.com/cramerlab/warp |
| PIXER[107] (Zhang et al., 2019) | CNN | Fully automated; Faster than DeepPicker and comparable with DeepEM in terms of time efficiency; The precision-recall performance is as good as DeepEM | Accuracy is lower on real-world datasets when compared to simulated data; Requires a dynamic updating strategy for the segmentation network | https://github.com/ZhangJingrong/PIXER |
| Topaz[108] (Bepler et al., 2019) | CNN & AE | Requires only a minimal amount of labeled data; Capable of picking challenging unusually shaped proteins; Can pair with autoencoder-based regularization to reduce the amount of labeled data; Uses a varied threshold | The precision-recall performance varied from ~30%-80% for 3 maps tested in the paper | http://cb.csail.mit.edu/cb/topaz |
| crYOLO[109] (Thorsten et al. 2019) | CNN | Dramatically reduced the runtime compared to previous DL approaches; The precision-recall performance is constantly higher than 90% in the test; | Vulnerable to template bias, a user needs to select particles for the training data that is representative to the full dataset | https://sphire.mpg.de/wiki/doku.php?id=pipeline:window:cryolo |

| | | Robust against low signal-to-noise (SNR) images | | |
| --- | --- | --- | --- | --- |
| AutoCryoPicker[111] (Al-Azzawi et al. 2019) | Intensity based clustering (IBC) | Fully automated; Outperformed two standard clustering algorithms (K-means and FCM) in both accuracy and runtime; The precision-recall performance is constantly higher than 95% in all tests, which is the highest amount of all current approaches | Detects only circle and square shape particle with Circular Hough Transform algorithm (CHT) | https://github.com /jianlin-cheng/Aut oCryoPicker |

## 2.2. 3D reconstruction

3D reconstruction in single-particle analysis uses algorithms to reconstruct a 3D map from 2D projection images. Cryo-EM single-particle analysis can involve purified proteins or nucleic acid complexes. Carbon film images ideally contain 2D copies of the macromolecular complex. 3D reconstruction is complicated by the unknown orientation of molecules in the 2D images and noisy projections. Addressing heterogeneity of the biological molecules is the main challenge for 3D reconstruction. An early method of 3D reconstruction used 3D DNA-Origami, a technique involving flexible and customizable designs of 3D structures with the use of 2D arrays of DNA scaffolding to bind in proteins for the cryo-EM sample[114]. The approach was tested on p53, a tumor suppressor which is a proline-rich region and a structured DNA-binding domain. The p53 structure has also been solved with X-ray crystallography and Nuclear Magnetic Resonance (NMR) as a comparison. Although the process used five different positions of the p53 binding sequence on a central double-stranded DNA (dsDNA) helix, there were incorrect attachments in the subset of structures, emphasizing the importance of refining the 2D averages of individual particle images. cryoDRGN is a neural network based approach that uses *ab initio* reconstruction of 3D protein complexes using simulated and real 2D cryo-EM image data to solve the 3D reconstruction[115]. The framework is trained to forward image formation from 3D views to a 3D volume, and the two strategies used to inference the system is a variational approach for a latent variable related to the protein's structural heterogeneity and a global search for unknown variations of each image. CryoSPARC is a platform that combines innovations in 3D reconstruction algorithms to provide a streamlined single-particle cryo-EM workflow[116]. Its software package is used to solve cryo-EM maps of membrane proteins, viruses, complexes, and small particles, with the process being efficient due to optimized algorithms and GPU accelerations of the pipeline.

# 3. AI in de novo molecular structure modeling from cryo-EM map

Over the last few years, the resolution of the cryo-EM maps has been greatly improved. There are different types of ML methods that have been applied to process a variety of density maps at low, medium, and high-resolutions, especially when homology structures were not available for rigid or flexible fitting. Many de novo structure modeling methods based on DL were also established in recent years. The methods vary with diverse capabilities in learning structural details, including secondary structure elements, Cα atoms, amino acid types, and side-chain poses. **Figure 2** depicts a comparison between the ML and DL approaches applied in cryo-EM modeling. **Table 2** also summarizes some recent achievements in ML and DL for protein structure modeling with cryo-EM maps.

## ML Methods vs DL Methods

| Factors | Machine Learning | Deep Learning |
|---|---|---|
| Analysis Goals | Solves low resolution modeling problems, focusing on secondary structures | Solves high resolution modeling problems, amino acids types, carbon alpha atom positions |
| Transparency | Easy to follow model's logic | Difficult to explain model's logic |
| Data Requirement | Low resolution is acceptable, less cryo-EM data required | Requires high resolution and lots of cryo-EM data |
| Hardware Requirements | Uses CPU and standard hardware | Requires GPUs and processing power |
| Training | Takes less time to train | Takes a longer time to train |
| Learning Techniques | K-Means, SVM, KNN | CNN, GCN, U-net |

**Figure 2**. A comparison with traditional ML and DL Methods with regards to protein structure modeling.

De novo modeling can be used when a 3D homology template is not available to predict the protein's tertiary structure, while amino acid sequence and cryo-EM map is available. Early de novo modeling focused on methods that did not involve cryo-EM imaging for a few reasons. First of all, cryo-EM generates noise from its procedure, and experimental data resulted in incomplete structures, requiring NMR or X-ray crystallography to complete the model. The breakthrough of image processing and direct electron detectors allowed cryo-EM to obtain resolutions that were better than 5 Å, which would allow cryo-EM to be used instead of relying on X-ray crystallography or NMR. Researchers could begin work with heterogeneous samples and visualizing macromolecular complexes due to the improved resolution[117].

The general process of de novo modeling from cryo-EM maps is to identify local dense voxels, then identify which points correspond to the backbone structure of a protein[118], followed by a sequence assignment procedure to build the structure model. Different models can be ranked in a score function and refined to determine the best models. Traditional de novo modeling has been limited to small proteins because its process requires huge computational resources and space. Since traditional de novo modeling uses image processing and physical laws based predictions, the algorithm might not be able to obtain an optimal solution in a very large complex[28]. Another limitation came from the poorly resolved density regions and heterogeneous data issues in cryo-EM maps. Examples of traditional de novo cryo-EM modeling methods include EM Fold[119], Gorgon[120], Rosetta[121], Pathwalking[28], MAINMAST[122], and Phenix[123].

Designing effective algorithms to fold 1D amino acid sequence into a specified 3D structure has attracted many researchers from the biotechnology and medical field because these structures can determine the functions of these proteins[124]. This past decade has seen multiple improvements in predicting and designing 3D protein structures. Software and automation advances have also allowed the growth of the biological database for experimentally determined protein structures which is now around 173,000 structures[125]. Additionally, both CPU and GPU increased computing power has allowed researchers to examine larger molecule and protein complex structures. With the growth of available data for training, researchers can now use ML algorithms to decipher the protein complexes. Over the past ten years, more and more artificial intelligence based modeling techniques have been developed due to improvement of imaging resolution and the increase of data (**Figure 3**).

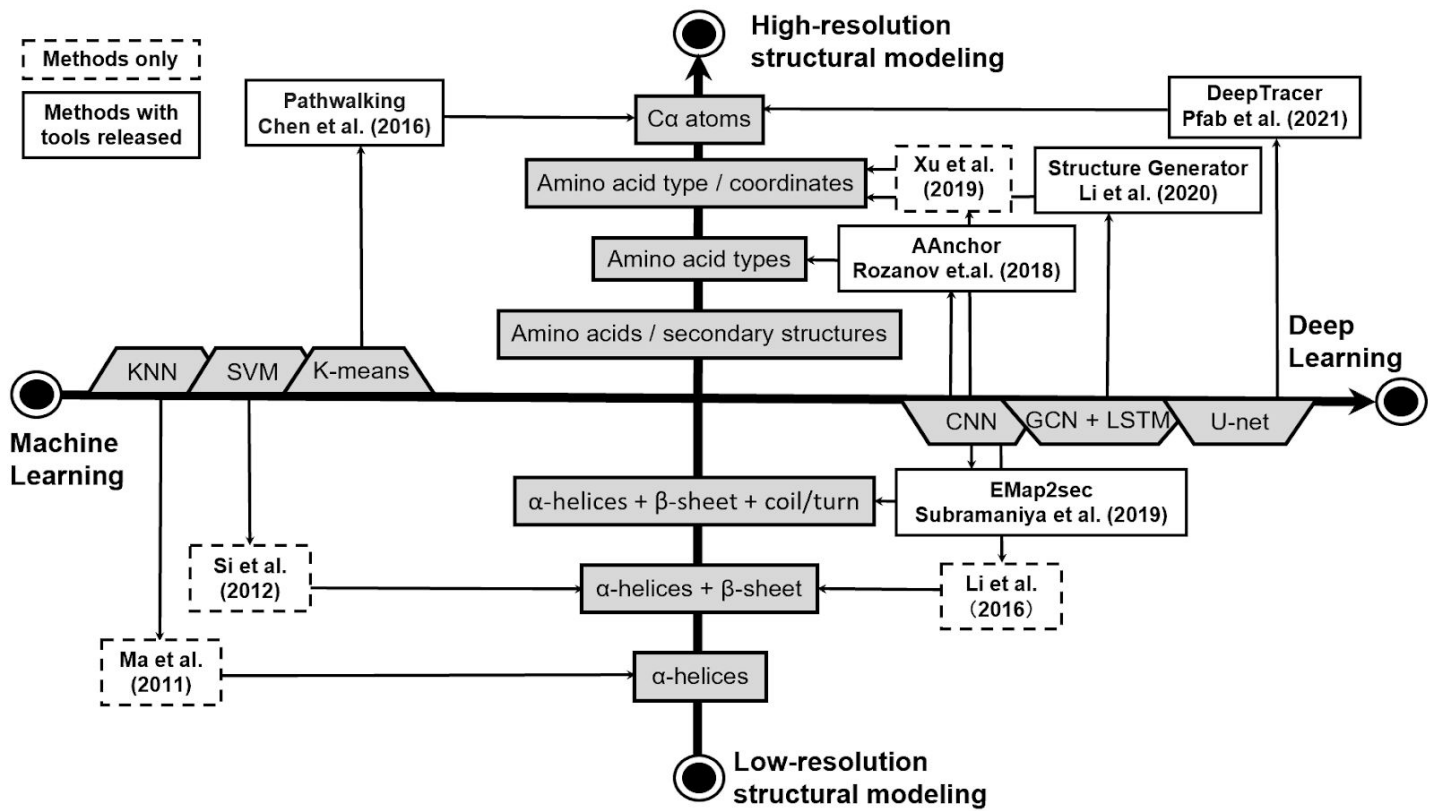

**Figure 3**. The growth of artificial intelligence based de novo methods with resolution for protein structure modeling.

Although nucleic acids have important roles in cellular activities and regulatory processes, fewer approaches have been proposed for 3D structural modeling of RNA/DNA interactions. There are less experimentally determined RNA/DNA structures available in databases. RNA/DNA structures are less conserved and cannot be reliably built through homology building[126]. Additionally, RNA structure modeling includes Watson-Crick interactions, atomic clashes, and ligand binding predictions, making it difficult to validate RNA structures[127]. Very few computational methods for predicting RNA secondary and tertiary structures are currently being developed.

**Table 2**. Strengths and limitations of ML/AI based de novo protein structure modeling methods.

| Method | Strengths | Limitations | Tool |
|---|---|---|---|
| RENNSH[23] (Ma et al., 2012) | Automated secondary structure detection; Tested on simulated maps of resolutions 6, 8, and 10 Å as well as experimental maps of resolutions 3.8, 6.8, and 8 Å, and performs the best amount of three previous methods. | Detects only α-helices location; Does not build an atomic model; Designed only for medium resolution ~5–10 Å | N/A |
| SSELearner[21] (Si et al., 2012) | Automated secondary structure detection; Tested on simulated maps at 8 Å and experimental maps of resolutions 3.8–9 Å | Detects only α-helices and β-sheets; Does not build an atomic model; Designed only for medium resolution ~5–10 Å | N/A |
| Pathwalking[28] (Chen et al., 2016) | Semi-automated backbone modeling; Incorporated SSE detection for improving the accuracy | Performance decreased on low-resolution maps that fall past 5 Å; Vulnerable to superfluous density or poorly resolved density; Algorithm may fail to converge to an optimal solution in the cases of large complexes | https://blake.bcm.edu/emanwiki/Pathwalker |
| A²-Net[27] (Xu et al., 2019) | Fully automated without manual tuning; High coverage and mean average precision; Fast in computation | KNN-graph selection is arbitrary | N/A |
| Structure Generator[26] (Li et al., 2020) | Fully automated pipeline; Improved location of Cα; Predicts amino acid even where Cα is missing; Correct mistakes from erroneous input | Work-in-progress, only tested on simulated maps; Prediction accuracy depends on input structural information; Amino acid sequences are limited to 700 residues | https://github.com/leeneil/structure-generator |
| Emap2sec[22] (Subramaniya et al., 2019) | Well performed on density maps of intermediate resolutions; Can apply to both simulated data and experimental data | Only secondary structure is resolved; Relatively poor performances on experimental data compared to simulated data | https://kiharalab.org/emsuites/emap2sec.php |
| AAnchor[101] (Rozanov and Wolfson, 2018) | Identify amino acid type and location with high confidence | Require high resolution density map (3.1 Å or better); Perform well only on amino acids with less rotamers; Overall identification fraction is low | http://bioinfo3d.cs.tau.ac.il/AAnchor |
| CNN by Li et al.[25] (Li et al., 2016 ) | Identify both helices and sheets with high sensitivity and specificity; Significantly better results than a SVM method SSELearner | Only simulated data validation | N/A |
| DeepTracer[19] (Pfab et al., 2021) | Fully automated; Can model atomic level protein features including backbone atom locations, secondary structures as well as amino acid types; High accuracy on Cα atoms modeling; Can model multi-chain large complex; Very fast in computation | Not perfect on amino acid assignments; Works better on maps of high resolution (<4 Å) | https://deeptracer.uw.edu |

## 3.1. Protein structure modeling using traditional machine learning techniques

The representative ML algorithms that were used to deal with cryo-EM data were K-mean clustering, K nearest neighbor (kNN), and Support Vector Machine (SVM). RENNSH[23] uses kNN for protein secondary structure identification. RENNSH is able to automatically detect α-helix based on the invariant properties in the voxels on the helix axis for medium resolutions density maps. The authors chose the spherical harmonic descriptor (SHD) to describe the local pattern of each voxel and kNN as the classifier to distinguish helix from non-helix voxels. The method was tested on simulated maps of resolutions 6, 8, and 10 Å as well as experimental maps of resolutions 3.8, 6.8, and 8 Å, and it performs the best amount of the three previous identification methods. SSELearner[21] is a ML approach that identifies α-helices and β-sheets in medium resolution maps through learning from existing volumetric maps in the Electron Microscopy Data Bank. It first uses a SVM as the classifier to classify α-helices, β-sheets, and background voxels based on the extracted features. Then, it determines the exact position for the secondary structures through post-processing. The method was tested on 10 simulated maps of resolutions 8 Å and 13 experimental maps of resolutions 3.8–9 Å. The resulting accuracy for α-helices detection for simulated maps and experimental maps was 91.8% and 74.5% respectively, and for β-sheets detection, accuracy is 85.2% and 86.5% respectively. Pathwalking[28] is a semi-automated de novo modeling approach that builds a protein Cα model from near-atomic resolution (3-6 Å) density maps using K-means clustering techniques and the traveling salesman problem (TSP) solver, and the steps are as follows.

## 3.2. Protein structure modeling using advanced deep learning techniques

With the advent of near-atomic resolution in cryo-EM imaging techniques and increased computational power, combined with the ever-larger dataset, DL has become one of the most popular algorithms used in cryo-EM structure modeling. Specifically, the CNN and its variations have shown their exceptional performances in pattern recognition from cryo-EM densities. Some other works also take advantage of recurrent neural networks and graph neural networks. In this section, we first review the application of DL methods in secondary structure prediction, followed by the prediction of atom coordinates and amino acid types from cryo-EM.

Emap2sec is able to identify protein secondary structures from cryo-EM maps with intermediate resolutions (5 – 10 Å)[22]. Emap2sec takes advantage of a two-phase convolutional neural network that combines local structural information with nearby regional features for the identification. In phase 1, Emap2sec takes an $11 \times 11 \times 11$ $Å^3$ voxel into a convolutional neural network to output the probability of the voxel involved in α-helix, β-sheet, or other structures. In phase 2, the probability values in $3 \times 3 \times 3$ $Å^3$ voxels are fed into fully connected neural networks to yield the final decision of the secondary structure. Emap2sec can achieve ~80% accuracy on average for the simulated maps and 64% for the experimental data. Emap2sec showed much superior performances on secondary structure identification compared to Phenix[128] and ARP/wARP[129], which are traditional modeling software for density maps of high resolution. Emap2sec also yielded better results on structure accuracy on some similar methods like HelixHunter[24] and CNN by Li et al[25]. CNN by Li et al. is a CNN-based algorithm to detect protein secondary structures in cryo-EM maps with high sensitivity and specificity[25]. To distinguish and learn different features between α-helices and β-sheets, the algorithm includes both inception learning and residual learning in 6 modules. To improve the computation efficiency, the algorithm also uses several deconvolution layers to offset size reduction and establish an end-to-end mapping between inputs and outputs. The algorithm has an average sensitivity and specificity of 71.52% and 97.86% respectively on α-helix identification for simulated cryo-EM density images. For β-sheet detection, it achieves 76.04% and 91.87% for sensitivity and specificity. Compared to SSELearner[21], which utilizes a multi-class SVM algorithm for protein secondary structure identification, the CNN algorithm shows significantly better performance with an average F1 score of 73.08 compared to 50.62 for SSELearner on secondary structure identification.

Deep learning also demonstrated the effectiveness in the identification of amino acids with their types, location, and pose from the input cryo-EM map. $A^2$-Net is a two-stage protein structure modeling algorithm which uses DL and Monte Carlo search to determine protein structure from cryo-EM maps[27] . First, amino acid network ($A^2$-Net) utilizes three DL architectures to learn 20 individual amino acid type and their poses from 250K amino acids in 1713 protein chains along with their cryo-EM maps, including localization network (locNet), recognition network (recNet) and pose estimation network (poseNet). Next, a four-step Monte Carlo Tree Search (MCTS) algorithm is applied to connect each amino acid determined from $A^2$-Net into a protein chain. To save computation time, a KNN-graph is built to select proposals with the number of amino acids that match the ground truth amino acid sequence. A peptide bond recognition network that detects peptide bonds between two amino acids is also applied to further speed the Monte Carlo search. The optimal proposal is chosen as the protein structure after enough cycles of simulation. The $A^2$-Net is found to have higher mean average precision and coverage compared to 3D-VGG and 3D-ResNet. In terms of computation time, $A^2$-Net has much less computation time with only 11.3 minute compared to Rosetta-denovo (>100h) on the same protein.

Structure Generator is a modeling method that utilizes graph convolutional network (GCN) and recurrent network (RNN) to construct a 3D protein structure with individual amino acid information extracted from cryo-EM maps[26]. The input of Structure Generator is the amino acid types with their rotameric orientation. These identities are obtained by applying CNNs on the cryo-EM densities. In the work, a CNN called RotamerNet is used. With amino acids' identity and their Cα coordinates as nodes, a two-layer GCN that connects any two proposed Cα positions within 4 Å by an undirected edge is established to propagate messages. After that, Structure Generator utilizes a bidirectional LSTM to label candidate identities and atomic locations consistent with input protein sequences in order to construct the whole protein chain. The Structure Generator is able to get high prediction accuracy (>0.99) on ProteinNet[130] and UniRef50[131] datasets. However, the average accuracy with structural information from their own RotamerNet is moderate. Interestingly, although overall accuracy of Structure Generator is limited by the accuracy of RotamerNet, many proteins tested have higher Structure Generator accuracy than RotamerNet, which indicates Structure Generator can predict correct structure with some missing or false information. One major limitation for Structure Generator is that the mean sequence length used for validation was only 201.3 residues.

AAnchor is a ML program that can detect certain amino acids with high confidence in high-resolution cryo-EM maps (3.1 Å or better)[101]. AAnchor consists of two building blocks: classification and detection. The classification block is based on convolutional neural networks (CNNs) that label the input $11 \times 11 \times 11$ $Å^3$ voxel with the most probable amino acid. The input voxel is fed into three identical-structured CNNs that are trained on only experimental data ($N_E$), only simulated data($N_S$) and the mix of the two ($N_{ES}$) in parallel. The type and probability of an amino acid in the voxel is determined based on certain statistics ($N_E$, average, majority, etc.) of the results from the three CNNs. The detection block then outputs the center of mass of the amino acid with its type and level of confidence. AAnchor shows the best results on LEU, GLY, ALA, VAL, LYS, TYR and PRO, which don't have many rotamers. Interestingly, AAnchor yields significantly better results on density maps of 2.9 Å and 3.1 Å than on maps of 2.2 Å in which the authors argue to be a result of the differences in availability of the training data with different resolutions. The overall identification fraction of AAnchor is 10%-20%.

DeepTracer is a fast and fully automated software that can predict protein backbone atom positions, secondary structures, and amino acid types from protein cryo-EM maps using deep learning[19]. DeepTracer evolves from a previous work of cascaded convolutional neural network (C-CNN)[132]. Preprocessed density maps are first cut into $64 \times 64 \times 64$ voxel cubes and then fed to DeepTracer's neural network. Each voxel cube will go through four different U-Nets[133], namely Atoms U-Net, Backbone U-Net, Secondary Structure U-Net, and Amino Acid Type U-Net to extract different levels of structural features. With the output of Cα channel from Atoms U-Net, DeepTracer first finds out coordinates of Cα atoms by calculating the center of mass around the local maximums and then utilizes an optimized graph algorithm to connect individual Cα atoms into protein backbone. After helix refinement and some other post-processing steps, protein side-chain atoms can be plugged in using Scwrl[134]. DeepTracer can predict Cα locations in the protein backbone with high accuracy. On Phenix Benchmark Test with 476 cryo-EM maps, DeepTracer has an average of 76.93% compared to 45.65% with Phenix[135] on residue matching, which represents an improvement of over 30%. For sequence matching, DeepTracer also demonstrates significantly better results with 49.83% compared to 12.29% by using Phenix[135]. DeepTracer has exceptionally fast computation speed as well. It can finish tracing a large multi-chain complex with around 60000 residues in 2h[19].

## 3.3. RNA/DNA structure modeling approaches

Advances in RNA/DNA Modeling allow cryo-EM maps to model RNA/DNA in addition to proteins secondary structure elements. The three approaches are comparable with other state of the art methods. **Table 3** shows the strength and limitations of each RNA/DNA structure modeling method. Haruspex is a pre-trained network that works on cryo-EM maps to determine protein secondary structure and RNA/DNA mapping. With the neural network Haruspex trained on 293 cryo-EM maps, with a resolution less than 4 Å, the network's recall was 95.1% and precision was 80.3%, for 122 maps[136]. Their novel method uses ML to identify nucleotides in Cryo-EM maps. For limitations, Haruspex misclassifies semi-helical structures, β-hairpin turns, and residues belonging to polyproline are misclassified as α-helical, while loosely parallel structures are frequently misclassified as β-strands[137]. The combination of multi-dimensional chemical mapping and Rosetta DRRAFTER computational modeling allows for a hybrid 'Ribosolve' pipeline. In the pipeline, RNA structure is determined through cryo-EM maps and next-generation sequencing (M2-seq) with auto-DRRAFTER[138]. AutoRRAFTER models RNA coordinates into the cryo-EM maps, in order to confirm or refute homology between RNA classes to validate structure prediction for nanostructures. The application is specific for RNA modeling. It does not correct the position of atoms and cannot examine the accuracy of positions for RNA bases, base pairs, phosphates and metal ions[138]. ARP/wARP is a model building tool that obtains the PDB structure of a protein and protein-nucleic acid complex from a cryo-EM map[139]. Model building uses ML-based docking of main-chain fragments, and a library of protein fragments for sequence-independent connections[140]. The main-chain is built with plausible connections between Cα atoms. SVM classifiers are used to estimate residue-type probabilities and find out which alignment probabilities are for each fragment to the target sequence. ARP/wARP currently can model main-chain fragments from maps where the location resolution is 4 Å or better[139]. Currently, only the protein metrics are evaluated in the model.

**Table 3**. Strengths and limitations of RNA/DNA structure modeling

| Method | Strengths | Limitations | Tool |
|---|---|---|---|
| Haruspex[136] (Mostosi et al., 2020) | Accurate recall; Builds on ground truth data; Compatible with main-chain programs to map the primary and secondary structures | False positive readings with helices and sheets; Secondary Structures misclassified as α-helices and β-strands | https://github.com/thorn-lab/haruspex |
| Auto-DDRafter[138] (Kappel et al., 2020) | Accurate results with NMR; Uses multiple models for checking conformations of 3D structure | Does not account for the positioning of atoms; Specific for RNA structures | https://www.rosettacommons.org/demos/latest/public/auto-drrafter/README |
| ARP/wARP[139] (Chojnowski et al., 2021) | Accurate predictions for small side chains; Accurate at lower resolution | Training data if overfitted, can only sample a few models; Only accesses the protein in evaluation | https://arpwarp.embl-hamburg.de/ |

# 4. Outlook and Discussion

The creation of innovative tools and new technologies for AI-based cryo-EM modeling will pave the way for future study of macromolecular structure and function. The computational tools and sophisticated analysis approaches will lay a foundation for advances in drug discovery and modern molecular biomedicine, and will play a key role in preventive and precision diagnosis and treatment. With the rapid development of cryo-EM and artificial intelligence. AI-based cryo-EM modeling methods will make de novo approaches more intelligent and inexpensive.

## 4.1. AI for quality assessment in cryo-EM modeling

Currently there are several standard metrics for validating the quality of cryo-EM produced structure (e.g., the Fourier shell correlation (FSC), Molprobilty[141], EMRinger[142], CaBLAM[141]). However, none of them used ML techniques, which have been successfully used in the protein structure predictions to evaluate the quality of predicted structures (e.g., DeepQA[143], TopQA[144], AngularQA[145], ProQ3D[146]). The latest 2019 EMDataResource challenge has reported recommendations to be made about validating near-atomic cryo-EM structures both in the context of individual experiments and structure data archives[147]. ML has been used in this challenge, but there is still a long way to build the robust validation metrics for cryo-EM maps and models[148]. It is hard to detail visions of the future techniques for the best validation metrics, but it is certain that we will need more new techniques and tools for validating the structures in the cryo-EM field, and ML techniques have great potential for improvement in the cryo-EM field.

## 4.2. AI for structure dynamics and functions in cryo-EM

Elucidation of both the 3D structure and the dynamics of a protein is essential to understand its function. However, the analysis of the dynamics using cryo-EM is often challenging because of complexity of the structural assemblies. Recently, a DL-based approach has been developed to extract the atomic fluctuations that are hidden in cryo-EM maps and detect changes in dynamics that are associated with molecular recognition[149]. Macromolecules dynamics are also associated with transition states. Researchers have gained insight into the structural changes and transition by determining cryo-EM structures of Saccharomyces cerevisiae intermediates in the path from the 90S to the pre-40S[150]. Small molecules such as ligands, waters, and ions also play a key role in molecular interactions and functions. How reliable are ligands, waters, and ions built into cryo-EM maps and can they be placed automatically into the map are still questions faced by scientists and researchers in the field.

## 4.3. Other advanced AI methods for future cryo-EM modeling

To extract or learn features from cryo-EM maps for protein structure modeling, the most direct method in DL is CNN. There are already several successful works in protein secondary feature detection, backbone construction, and side-chain prediction with CNN. In the meantime, as protein is synthesized under the guidance of the amino acid sequence, algorithms that perform well with sequence inputs might also be effective in protein structure modeling. Other DL methods, which can also take advantage of the amino acid sequence of the protein, such as reinforcement learning (RL), generative adversarial networks[151], and transformers[152,153], will have great potential for better protein structure prediction as well[154]. DL techniques combining both information from cryo-EM images and amino acid sequence will be promising in more precise atomic-level protein structure modeling.

# Funding Information

This material is based upon work supported by the National Science Foundation under Grant No. 2030381 and the graduate research award of Computing and Software Systems division at University of Washington Bothell to D.S·· Any opinions, findings, and conclusions or recommendations expressed in this material are those of the authors and do not necessarily reflect the views of the National Science Foundation.